\documentclass{INTERSPEECH2023}

\usepackage{amsmath,graphicx}
\usepackage{hyperref}
\usepackage{balance}
\usepackage{amsmath,amsfonts}
\usepackage{algorithmic}
\usepackage{algorithm}
\usepackage{array}
\usepackage[caption=false,font=normalsize,labelfont=sf,textfont=sf]{subfig}
\usepackage{textcomp}
\usepackage{stfloats}
\usepackage{url}
\usepackage{verbatim}
\usepackage{graphicx}
\usepackage{cite}
\usepackage{multirow}
\usepackage{multicol}
\usepackage{makecell}
\usepackage{diagbox}
\usepackage{color,soul}
\usepackage{ctable} 
\usepackage{flushend} 

\DeclareMathOperator*{\argmax}{arg\,max}

\interspeechcameraready

\title{Time-Domain Speech Enhancement for Robust Automatic Speech Recognition}
\name{Yufeng Yang$^1$, Ashutosh Pandey$^{1}$, and DeLiang Wang$^{1, 2}$}
\address{
  $^1$Department of Computer Science and Engineering, The Ohio State University, USA\\
$^2$Center for Cognitive and Brain Sciences, The Ohio State University, USA}
\email{\{yang.5662, pandey.99, wang.77\}@osu.edu}

\begin{document}
\maketitle
\begin{abstract}
It has been shown that the intelligibility of noisy speech can be improved by speech enhancement algorithms. However, speech enhancement has not been established as an effective frontend for robust automatic speech recognition (ASR) in noisy conditions compared to an ASR model trained on noisy speech directly. The divide between speech enhancement and ASR impedes the progress of robust ASR systems especially as speech enhancement has made big strides in recent years. In this work, we focus on eliminating this divide with an ARN (attentive recurrent network) based time-domain enhancement model. The proposed system fully decouples speech enhancement and an acoustic model trained only on clean speech. Results on the CHiME-2 corpus show that ARN enhanced speech translates to improved ASR results. The proposed system achieves $6.28\%$ average word error rate, outperforming the previous best by $19.3\%$ relatively.

\end{abstract}

\section{Introduction}
\label{sec:intro}

In real environments, acoustic interference is ubiquitous in speech communication, and negatively impacts the performance of speech-based applications such as smart home devices \cite{heymann2018performance} and conference transcription systems \cite{fu2021aishell}. To attenuate background noise, speech enhancement algorithms have been developed to estimate clean speech from noisy speech. These algorithms have achieved remarkable success in improving the quality and intelligibility of noisy speech \cite{wang2018supervised}. However, a major disappointment is that monaurally enhanced speech does not translate to improved automatic speech recognition (ASR), and this has been attributed to the distortion introduced by monaural speech enhancement algorithms \cite{wang_bridging_2019}. The divide between speech enhancement and ASR has persisted despite considerable research over decades to bridge enhancement and ASR \cite{cooke2001robust, raj2004reconstruction, narayanan2014investigation}. This study represents a new effort to bridge the fields of speech enhancement and ASR.

For monaural speech enhancement, we employ the recently proposed attentive recurrent network (ARN), which performs speech enhancement in the time domain \cite{pandey2022self}. Time-domain enhancement operates differently from spectral methods \cite{wang2018supervised, wang2020complex}. It directly predicts clean speech samples from noisy speech samples, where speech magnitude and phase are enhanced simultaneously \cite{fu2017raw}. ARN incorporates a recurrent neural network (RNN) and a self-attention mechanism, and produces excellent enhancement performance \cite{pandey2022self}. We employ ARN as the frontend of ASR.

In terms of robust ASR, prevailing approaches perform acoustic modeling directly on noisy speech for noise-dependent or noise-independent models, which are proven to be effective on CHiME-2 \cite{wang_bridging_2019}, CHiME-4 \cite{yang2022conformer}, and CHiME-6 \cite{chan2021speechstew} corpora. The drawback of such approaches is that noise-dependent models do not generalize well to untrained noises and noise-independent models need an enormous amount of noisy speech for training, which is not only costly but also infeasible in many real applications. An ASR model trained on noisy speech also results in an unavoidable performance gap compared with a corresponding model trained on clean speech, when tested on clean speech. To bridge the divide between speech enhancement and ASR, attempts have been made to perform acoustic modeling on enhanced speech \cite{wang_bridging_2019, meng2018adversarial} or enhanced features \cite{wang_enhanced_2019} in a distortion-independent way. Nonetheless, backends of these systems are fully dependent on the frontend. When the frontend is replaced, say by a better enhancement model, the backend has to be retrained or ASR performance degrades. Likewise, speech enhancement can also be designed to serve ASR. In \cite{plantinga2021perceptual}, a perceptual loss based model was proposed to guide the training of a speech enhancement frontend using senone labels from an acoustic model. To completely combine speech enhancement and ASR, a speech enhancement model and an acoustic model can be jointly trained \cite{wang2016joint, zhu2022joint}. Similar ideas are also utilized in end-to-end (E2E) systems \cite{shi2022train, chang2022end}. Such E2E systems often have a huge model size and are difficult to train. Furthermore, the frontend and backend inside a joint or E2E system are dependent on each other, which makes it problematic to improve either the frontend or the backend individually.

In this paper, we investigate the most straightforward approach to robust ASR where the acoustic model is based on only clean speech using a Conformer-based acoustic model \cite{yang2022conformer} and its input is ARN enhanced speech directly. Thus the proposed robust ASR system completely decouples the frontend and backend. In other words, the frontend is designed for speech enhancement only, and the backend acoustic model for recognizing clean speech only. Combining two modules directly to recognize noisy speech, the proposed system is demonstrated to outperform other robust ASR systems, including noise-independent, noise-dependent, and distortion-independent models. When tested on the medium vocabulary track (track 2) of the CHiME-2 corpus, our best system achieves $6.28\%$ average word error rate (WER). To our knowledge, this WER result represents the best on this dataset to date and outperforms the previous best by $19.3\%$ relatively. Our investigation also shows that using short-time objective intelligibility (STOI) \cite{taal2011algorithm} as the model selection criterion is superior for speech enhancement models in terms of ASR. ARN enhanced speech also shows consistent improvements on other baseline acoustic models, which further demonstrates that ARN eliminates the divide between speech enhancement and ASR.

The remainder of the paper is organized as follows. Section~\ref{sec:system} describes ARN for time-domain enhancement and a Conformer-based acoustic model. Section~\ref{sec:exp} describes the experimental setup and implementation details. Evaluation results and comparisons are presented in Section~\ref{sec:result}, and Section~\ref{sec:conclusion} concludes the paper.

\section{System Description}
\label{sec:system}
\subsection{Problem Formulation}
The monaural speech enhancement problem is formulated as follows

\begin{equation}\label{eq:se}
\begin{aligned}
\mathbf{y} = \mathbf{s} + \mathbf{n},
\end{aligned}
\end{equation}
where $\mathbf{y}$, $\mathbf{s}$, and $\mathbf{n}$ are noisy speech, clean speech, and additive noise, respectively. Speech enhancement computes an estimate of $\mathbf{s}$, $\hat{\mathbf{s}}$, from $\mathbf{y}$. 

An ASR system computes the optimal word sequence $\mathbf{W^{*}}$ given a sequence of acoustic features $\mathbf{X}$ of speech signal $\mathbf{x}$, which is formulated as a maximum \emph{a posteriori} probability problem

\begin{equation}\label{eq:map}
    \mathbf{W^{*}} = \argmax_{\mathbf{W}} P_{\mathcal{AM}, \mathcal{LM}}(\mathbf{W} | \mathbf{X}),
\end{equation}
where $\mathcal{AM}$ and $\mathcal{LM}$ denote an acoustic model (AM) and language model (LM), respectively. Using Bayes' theorem, Eq.~\ref{eq:map} can be rewritten as

\begin{equation}
    \mathbf{W^{*}} = \argmax_{\mathbf{W}} p_{\mathcal{AM}}(\mathbf{X} | \mathbf{W})P_{\mathcal{LM}}(\mathbf{W}),
\end{equation}
where $p_{\mathcal{AM}}$ and $P_{\mathcal{LM}}$ are AM likelihood and LM prior probability, respectively. An AM predicts the likelihood of acoustic features of a phoneme or another linguistic unit, and an LM provides a probability distribution over words or sequences of words in a speech corpus.

\subsection{Attentive Recurrent Network}
\label{ssec:arn}
We employ ARN as the frontend of ASR, which is a time-domain speech enhancement model comprising RNN, self-attention, feedforward network, and layer normalization modules. Details of ARN building blocks can be found in \cite{pandey2022self}. In this work,  the non-causal version of ARN is used, namely the RNN in ARN is bi-directional long short-term memory (BLSTM) and self-attention is unmasked.



\begin{figure}[htbp!]
    \centering
    \includegraphics[width=0.941\linewidth]{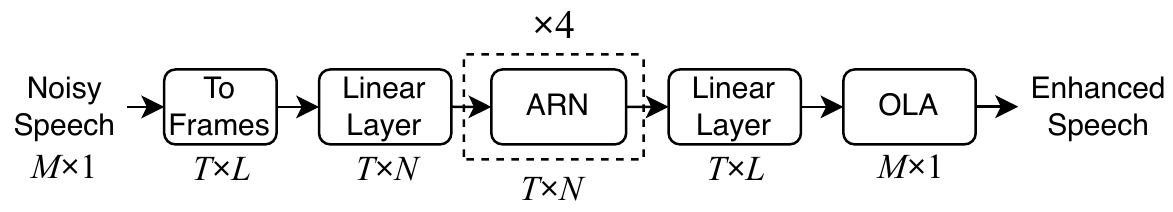}
    \caption{Diagram of ARN. $M$, $L$, and $T$ denote the number of samples of an input signal, frame length, and the total number of frames, respectively.}
    \label{fig:arn}
\end{figure}

The diagram of ARN is shown in Fig.~\ref{fig:arn}. After an input signal is chunked into overlapping frames, all frames are projected into a latent representation of size $N$ by a linear layer. Then the representations are processed by four consecutive ARN blocks. Finally, another linear layer projects the output of the last ARN block back to size $L$. The enhanced speech is finally computed using the overlap-and-add (OLA) method.

\subsection{Conformer-based Acoustic Model}
\label{ssec:conformer}

\begin{figure}[htbp!]
    \centering
    \includegraphics[width=0.9\linewidth]{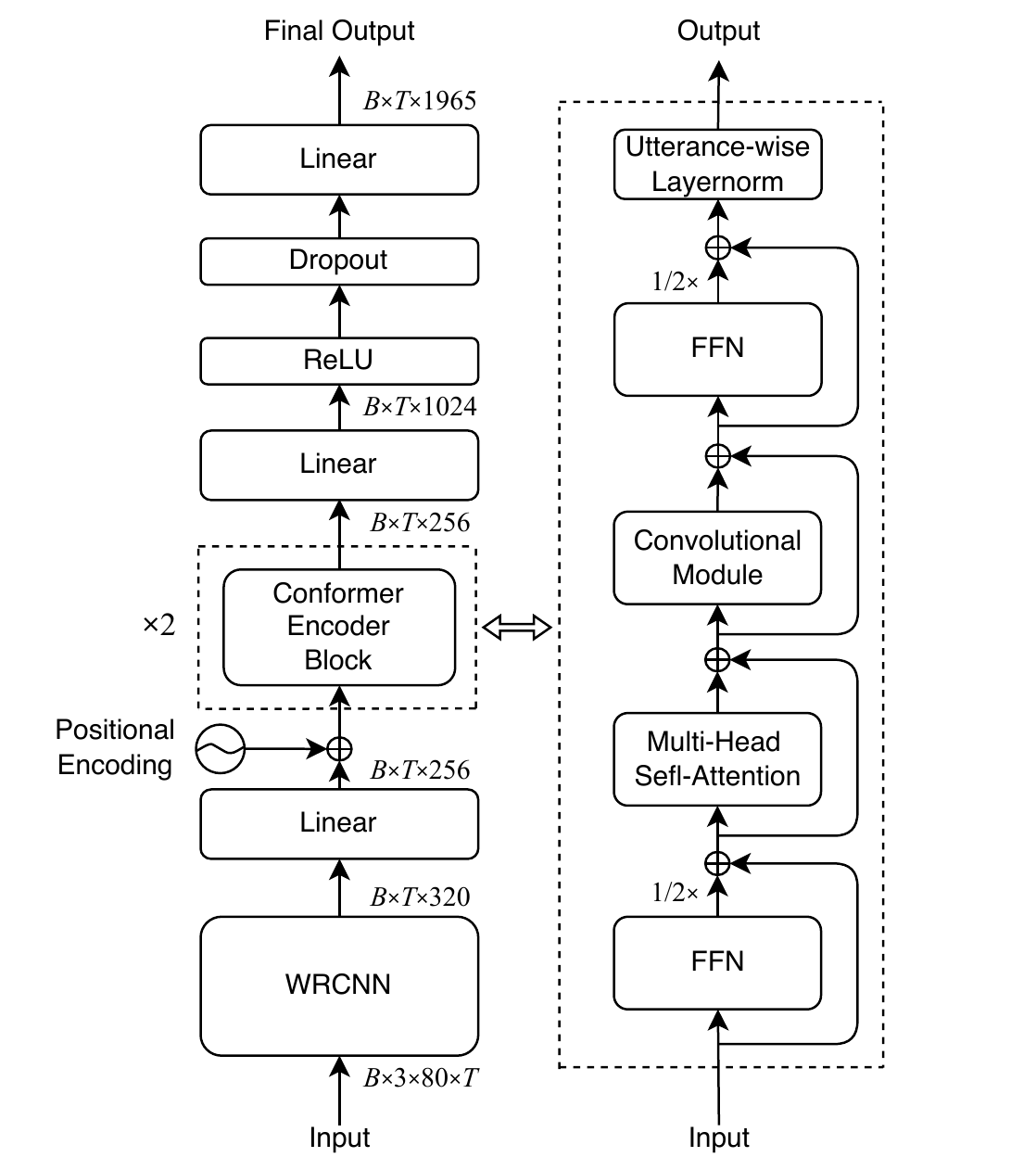}
    \caption{System architecture of Conformer-based AM. $B$ denotes the batch size, and $T$ denotes the number of time frames of the longest utterance in a batch.}
    \label{fig:cam}
\end{figure}

We utilize the Conformer-based AM \cite{yang2022conformer} as the backend in the proposed system. It is built upon a wide residual BLSTM network (WRBN) (see \cite{wang_bridging_2019}). The Conformer-based AM is shown to outperform WRBN on the CHiME-4 single-channel track \cite{yang2022conformer}. The system architecture of the Conformer-based AM is shown in Fig.~\ref{fig:cam}, where FFN denotes a feedforward network.

The Conformer-based AM takes as input 80-dimensional mean-normalized log-Mel filterbank features extracted from the ARN output (or other forms of output), coupled with its delta and delta-delta features. First, the input is processed by a wide residual convolutional layer denoted as WRCNN, which passes the input signal through a convolution layer and uses three residual blocks to extract representations at different frequency resolutions \cite{zagoruyko_wide_2016}. Afterwards, an utterance-wise batch normalization and a linear layer with exponential linear unit (ELU) non-linearity are utilized to project the signal onto 320 dimensions. Then a linear layer projects the signal onto the dimension of multi-head self-attention. After two blocks of the Conformer encoder with absolute positional encoding, the signal is projected onto 1024 dimensions, followed by a ReLU (rectified linear unit) activation and dropout. Finally, a linear layer projects the signal to the final output of each frame as the posterior probability for 1965 context-dependent senone states. Then the output is sent to a decoder for final text transcripts.

\section{Experimental Setup}
\label{sec:exp}

\subsection{Dataset}
Our experiments are conducted on a medium vocabulary track (track 2) of the CHiME-2 corpus \cite{vincent2013second}. It is a commonly used dataset for robust ASR evaluation, which is generated by convolving Wall Street Journal (WSJ) clean speech with binaural room impulse responses (BRIRs) and mixing with non-stationary family home noise \cite{barker2013pascal}. Although the speech materials are from the WSJ corpus, due to the BRIRs during the data generation process \cite{christensen2010chime}, the speech alignments are altered, which makes it infeasible to use WSJ anechoic-clean speech as the training target for speech enhancement. Therefore, we treat reverberant-clean speech as the training target for ARN. Two channels are averaged to produce single-channel speech.

Training data for ARN is generated by randomly mixing 7138 reverberant-clean training utterances (`reverberated') from CHiME-2 with noise segments randomly picked from 10k non-speech sounds from a sound effect library (\href{http://www.sound-ideas.com}{http://www.sound-ideas.com}) under signal-to-noise ratio (SNR) uniformly distributed in [$-7$, $0$] dB and [$0$, $10$] dB ranges, with $50\%$ probability for each range to be selected. Validation data is generated by mixing 409 reverberant-clean validation utterances (`reverberated') from CHiME-2 with factory noise from the NOISEX-92 dataset \cite{varga1993assessment} under $-6$ dB SNR. Once trained, ARN is tested on reverberant-noisy test data (`isolated'), which has six SNR levels with each containing $330$ utterances.

The default backend in the proposed system is trained and validated on 7138 and 409 reverberant-clean utterances (`reverberated') from CHiME-2, respectively. We also train a noise-dependent backend (i.e. dependent on the CHiME-2 noise) which is trained and validated on 7138 and 409 reverberant-noisy utterances (`isolated'), respectively. Training data for the noise-independent backend is generated by randomly mixing 7138 reverberant-clean training utterances (`reverberated') with randomly picked 10k noise under the same SNR range as for ARN training. Training data for the distortion-independent backend is generated by enhancing the training data for the noise-independent backend using ARN. The noise-independent 
 and distortion-independent backends are trained on 157036 utterances in the same way as \cite{wang_bridging_2019}.

\subsection{Implementation Details}
The sampling rate for all utterances is 16 kHz. All training samples are generated randomly and dynamically for ARN. We apply root mean square normalization to noisy mixtures, and clean speech is scaled to produce a specified SNR level. During training, the number of samples for each utterance is set to 64000. Input and output frame size is set to 16 ms with a 2 ms frame shift. Dimension $N$ for BLSTM is set to 1024. The dropout rate is set to $0.05$ in feedforward blocks. ARN is trained using the PCM (phase-constrained magnitude) loss \cite{pandey2021dense}, which computes loss for both enhanced speech and noise in the time-frequency domain. Training over 157036 utterances is considered as one epoch, and ARN is trained for 100 epochs with batch size 16. The Adam optimizer \cite{kingma2015adam} is utilized. The learning rate of the first 33 epochs is fixed to $2e^{-4}$ and then exponentially decays every epoch till the final learning rate of $2e^{-5}$. Because STOI is shown to relate to WER \cite{moore2017speech}, thus we also use validation STOI as a model selection criterion in addition to validation PCM loss. 


For the backend, the configuration of WRCNN is kept the same as in \cite{yang2022conformer}. The attention dimension is set to $256$. The kernel size of the $1$-D depthwise convolution is set to $16$. We use a learning rate schedule from \cite{vaswani2017attention}, with $2$k warm-up steps and a learning rate factor $k$ of $100$. The Adam optimizer with $\beta_{1}=0.9$, $\beta_{2}=0.98$, and $\epsilon=1e^{-9}$ is utilized for model training. The batch size is set to $3$ and short utterances are padded with zeros to match the length of the longest utterance in each batch. The dropout rate is set to $0.15$ for network and attention weights. All backends are trained for $25$ epochs and model selection is based on the cross-entropy loss on the validation set. Training labels are 1965 senones in our experiments, and they are generated as in \cite{wang2016joint}. For log-Mel feature extraction, the approach in \cite{wang_bridging_2019} is applied, and pre-emphasizing, dithering, and direct current offset removal steps are skipped. A Hamming window is applied to the input waveform for STFT. Then a small value of $e^{-40}$ is added to prevent the underflow of logarithmic operation. For the noise-dependent, noise-independent, and distortion-independent backends, factor $k$ and the number of warm-up steps pairs are empirically set to \{$1e^{4}$, $5$k\}, \{$5e^{3}$, $80$k\}, and \{$5e^{3}$, $80$k\} respectively for better convergence and validation cross-entropy. Two distortion-independent backends are trained on the enhanced speech by ARN with minimum PCM validation and maximum STOI validation, denoted as distortion-independent PCM backend and STOI backend, respectively.

The decoder used in this work is the same as in \cite{wang2016joint}. AM outputs are first substracted by the log priors, and then fed to the decoder, which is based on the CMU pronunciation dictionary and the official 5k close-vocabulary tri-gram language model. The decoding beamwidth is set to 13, and lattice beamwidth is 8. The number of active tokens ranges from 200 to 700. Language model weights ranging from 4 to 25 are utilized.

\section{Evaluation and Comparison Results}
\label{sec:result}
\subsection{Results on Speech Enhancement}
Table~\ref{tab:se_chime2_dnarn} presents the speech enhancement results of ARN on the CHiME-2 corpus in terms of STOI \cite{taal2011algorithm} and perceptual evaluation of speech quality (PESQ) \cite{rix2001pesq}.  ARN with minimum validation PCM loss and ARN with maximum validation STOI are both evaluated. Under $-6$ dB SNR, ARN with minimum validation PCM and ARN with maximum validation STOI improve STOI by $19.64\%$ and $19.57\%$, and PESQ by $1.27$ and $1.29$, respectively. ARN with minimum validation PCM outperforms ARN with maximum validation STOI at all SNR levels except for PESQ under -6 dB SNR, although the performance differences between the two models are small. Enhancement results on CHiME-2 from other baseline models are not available, but the enhancement comparisons between ARN and other models are available in \cite{pandey2022self}.

\begin{table*}[htbp!]
    \centering
    \caption{Speech enhancement results of different ARN models on CHiME-2. STOI results are in percentage.}
    \label{tab:se_chime2_dnarn}
    \centering
    \scalebox{0.74}{
    \begin{tabular}[width=\linewidth]{| c | c | c | c | c | c | c | c | c|}
         \hline
        \multicolumn{2}{|c|}{Test SNR} & -6 dB & -3 dB & 0 dB & 3 dB & 6 dB & 9 dB & Avg. \\
         \hline
         \hline
         \multirow{3}{*}{\rotatebox{90}{STOI}} & Mixture & 73.68 & 77.80  & 81.34  & 85.16& 88.12  & 90.93 & 82.84 \\
         \cline{2-9}
         & \makecell[c]{Min PCM ARN} & \textbf{93.32}  & \textbf{94.44}  & \textbf{95.17}  & \textbf{96.01}   & \textbf{96.50}  & \textbf{97.13} & \textbf{95.43} \\
          & \makecell[c]{Max STOI ARN} & 93.25 & 94.17  & 94.71  & 95.35    & 95.75   & 96.24  & 94.91\\
          \hline
          \hline
          \multirow{3}{*}{\rotatebox{90}{PESQ}} & Mixture  & 2.13 & 2.33  & 2.49  & 2.66   & 2.85  & 3.04 & 2.58 \\
         \cline{2-9}
         & \makecell[c]{Min PCM ARN} & 3.40 & \textbf{3.50}  & \textbf{3.59}  &\textbf{3.68}   & \textbf{3.75}  & \textbf{3.81} & \textbf{3.62} \\
          & \makecell[c]{Max STOI ARN} & \textbf{3.42} & \textbf{3.50}  & 3.56  & 3.64   & 3.70  & 3.76 & 3.60\\
          \hline
    \end{tabular}}
\end{table*}

\subsection{Results of Acoustic Models Trained on Noisy Speech}
Evaluations of ASR performance are presented in Table~\ref{tab:asr_chime2_dnarn}, where \emph{Enhancement Model} denotes the frontend enhancement model of the system, and \emph{AM Type} denotes the type of training data for the backend. When tested on CHiME-2 reverberant-clean test utterances (`scaled'), the WERs for the default backend, noise-dependent backend, noise-independent backend, distortion-independent PCM backend, and distortion-independent STOI backend are $2.90\%$, $4.35\%$, $3.48\%$, $4.62\%$, and $4.33\%$, respectively.

\begin{table*}[htbp!]
    \centering
    \caption{ASR ($\%$WER) results of the proposed system and comparison systems on CHiME-2. Enh. denotes `enhanced'.}
    \label{tab:asr_chime2_dnarn}
    \centering
    \scalebox{0.76}{
    \begin{tabular}[width=\linewidth]{ | c | c | c | c | c|c|c|c|c|c |}
        \hline
         \multirow{2}{*}{Model Name} & \multirow{2}{*}{\makecell[c]{Enhancement\\ Model}} & \multirow{2}{*}{\makecell[c]{AM\\ Type}} & \multicolumn{6}{c|}{SNR} & \multirow{2}{*}{Avg.} \\
         \cline{4-9}
         & & & -6 dB & -3 dB & 0 dB & 3 dB & 6 dB & 9 dB & \\
         \hline
         \hline
         Mixture& - & Clean & 73.81 & 64.88  & 57.59 & 45.10 & 36.05 & 28.54 & 51.00 \\
         \hline
         Noise-independent WRBN \cite{wang_bridging_2019} & - & \multirow{4}{*}{Noisy} & 17.45 & 13.06 & 10.69 & 8.82 & 7.72  & 6.63 & 10.73 \\
         Noise-independent backend & - &  & 19.82 & 13.30 & 10.91 & 9.25 & 7.25 & 6.63 & 11.19\\
         Noise-dependent WRBN \cite{wang_bridging_2019} & - &  & 14.83 & \textbf{9.98} & 8.95 & 6.78 & 6.26 & \textbf{5.49} & 8.72 \\
         Noise-dependent backend & - &  & \textbf{14.44} & 10.33 & \textbf{7.92} & \textbf{6.73} & \textbf{6.03} & \textbf{5.49} & \textbf{8.49}\\
         \specialrule{.1em}{0em}{0em}
         Distortion-independent WRBN \cite{wang_bridging_2019} & GRN & \makecell[c]{Enh. Speech}  & 15.45 & 11.04 & 9.70  & 7.10 & 6.54 & 5.51 & 9.22\\
         \cline{3-3}
         Distortion-independent WRBN \cite{wang_enhanced_2019} & GRN & Enh. Feature  & 13.11 & 9.43 & 7.92 & 6.20 & 5.45 & 4.54 & 7.78\\
         \cline{3-3}
         Perceptual loss based \cite{plantinga2021perceptual} & Wide ResNet & \multirow{3}{*}{Clean} & 15.2 & 10.9 & 8.3 & 6.7 & 5.8 & 5.2 &8.7  \\
         Proposed & Min PCM ARN &  & 13.30 & 9.66 & 7.83 & 6.41 & 5.25 & 4.60 & 7.84 \\
         Proposed & Max STOI ARN &  & \textbf{9.94} & \textbf{7.04} & \textbf{6.50} & \textbf{5.45} & \textbf{4.54} & \textbf{4.22} & \textbf{6.28}\\
        \specialrule{.1em}{0em}{0em}
         \multirow{2}{*}{Noise-independent backend} & Min PCM ARN & \multirow{4}{*}{Noisy} & 12.93 & 9.14 & 8.28 & 6.65 & 5.53  & 4.86 & 7.90\\
         & Max STOI ARN &   & \textbf{9.62} & \textbf{7.34} & \textbf{6.78} & \textbf{5.45} & \textbf{4.84} & \textbf{4.46} & \textbf{6.42} \\
        \cline{1-1}
         \multirow{2}{*}{Noise-dependent backend} & Min PCM ARN &   & 12.54 & 9.25 & 8.29 & 6.73 & 6.15 & 5.87 & 8.14 \\
          & Max STOI ARN & & 9.73 & 7.40 & 7.23 & 6.41 & 5.16 & 5.23 & 6.86 \\
         \specialrule{.1em}{0em}{0em}
         \multirow{2}{*}{Distortion-independent PCM backend} & Min PCM ARN & \multirow{4}{*}{Enh. Speech}  & 10.14 & 7.98  & 7.29 & 6.28 & 5.94 & 5.51 & 7.19 \\
          & Max STOI ARN & & 8.84 & 6.89 & 6.50 & 6.13 & 5.57 & 5.29  & 6.54 \\
          \cline{1-1}
          \multirow{2}{*}{Distortion-independent STOI backend}& Min PCM ARN &  & 10.61 & 8.22 & 7.36 & 6.35 & 5.98 & 5.32 & 7.31 \\
          & Max STOI ARN & & \textbf{8.05} & \textbf{6.63} & \textbf{6.39} & \textbf{5.87} & \textbf{5.21} & \textbf{4.78} & \textbf{6.16} \\
         \hline
    \end{tabular}}
\end{table*}

We first evaluate and compare AMs trained on noisy speech. The noise-independent and noise-dependent WRBN in Table~\ref{tab:asr_chime2_dnarn} are trained on CHiME-2 clean speech mixed with 10k noise and CHiME-2 noisy speech, respectively \cite{wang_bridging_2019}. Compared with noise-independent WRBN, our noise-independent backend achieves reasonably good performance. The noise-dependent backend outperforms the noise-dependent WRBN by $2.6\%$, and is the best-performing model. The results demonstrate the effectiveness of the Conformer-based AM baselines.

\subsection{Results of the Proposed System}
We next evaluate the proposed system and compare it with other systems that incorporate speech enhancement. One system is WRBN trained on GRN (gated residual network) \cite{tan2018gated} enhanced speech \cite{wang_bridging_2019}, and the other is WRBN trained on GRN enhanced magnitude spectra \cite{wang_enhanced_2019}. A perceptual loss based model \cite{plantinga2021perceptual} is also included, and it employs a Wide ResNet trained with a perceptual loss as the frontend and its backend uses an off-the-shelf Kaldi CHiME-2 recipe. The proposed system with the ARN selected using minimum validation PCM achieves $7.84\%$ WER on average, which is close to the previous best $7.78\%$ from distortion-independent WRBN with enhanced features \cite{wang_enhanced_2019}. The proposed system with the ARN selected using maximum validation STOI  achieves $6.28\%$ WER, which outperforms the previous best by $19.3\%$ relatively. Switching the model selection criterion for ARN reduces WER from $7.84\%$ to $6.28\%$, corresponding to a relative improvement of $19.9\%$. This suggests that maximum validation STOI is a strong option for speech enhancement model selection when tested on downstream ASR tasks. Results from the proposed system demonstrate that speech enhancement advances translate to improved ASR results. In this way, the divide between speech enhancement and ASR is eliminated by the use of ARN speech enhancement.


\subsection{Results of ARN on Other Acoustic Models}
We finally test ARN enhanced speech on the noise-independent, noise-dependent, and distortion-independent backends for comprehensive comparisons. The WER results are given in the bottom parts of Table~\ref{tab:asr_chime2_dnarn}. Tested on ARN enhanced speech, WERs of noise-independent and noise-dependent backends are consistently reduced in comparison to those tested on noisy speech directly. With ARN selected using maximum validation STOI, the WER results for the noise-independent and noise-dependent backends are improved by $42.6\%$ and $19.2\%$ respectively over the corresponding results on noisy speech (see Table~\ref{tab:asr_chime2_dnarn}). The improvements show that ARN enhanced speech can benefit not only AM trained on clean speech only but also AM trained on noisy speech. With ARN enhanced speech as the input, even though ASR performances get improved for AMs trained on noisy speech, the proposed system with AM trained on clean speech still outperforms both baselines. Compared with distortion-independent backends, the proposed system achieves comparable performance with the $6.16\%$ WER from STOI backend tested on matched ARN enhanced speech, and outperforms at SNR higher than $0$ dB. This further demonstrates the utility of the proposed system that fully decouples the enhancement frontend and the backend trained on clean speech only.

\section{Concluding Remarks}
\label{sec:conclusion}
This study aims to eliminate the divide between speech enhancement and ASR. The time-domain enhancement model of ARN is employed as the frontend to a Conformer-based AM trained on clean speech only.  The proposed system fully decouples speech enhancement and ASR. Results on the CHiME-2 corpus show that better speech enhancement translates to improved ASR results. The proposed system achieves $6.28\%$ WER on the CHiME-2 corpus, outperforming the previous best by $19.3\%$ relatively. Future work includes conducting more experiments on other corpora, applying ARN to other ASR tasks such as continuous speech separation, and extending to multi-channel robust ASR tasks.

\section{Acknowledgements}
This research was supported by an NIH grant (R01DC012048), the Ohio Supercomputer Center, and the Pittsburgh Supercomputer Center (NSF ACI-1928147).


\bibliographystyle{IEEEtran}
\bibliography{mybib}

\end{document}